%% file: main.tex
\documentclass[a4paper,11pt]{article}
\usepackage{pos}

\usepackage{graphicx}\graphicspath{{figs/}}
\definecolor{darkblue}{rgb}{0,0,0.5}
\definecolor{firebrick}{rgb}{0.75,0.125,0.125}
\definecolor{darkgreen}{rgb}{0,0.5,0}
\usepackage[capitalize]{cleveref}
\usepackage{float}

\usepackage{wrapfig}
\usepackage{enumitem}

\newcommand{\eq}[1]{\begin{align}#1\end{align}}

\usepackage{lineno}

\title{Cloud Detection using Night Sky Background Light at the Pierre Auger Observatory}
 \ShortTitle{Night Sky Background Light for Cloud Detection}

\author*[a]{Fedor Tairli}

\affiliation[a]{School of Physics, Chemistry \& Earth Sciences, University of Adelaide, Adelaide, Australia}

\onbehalf{for the Pierre Auger Collaboration$^b$}
\affiliation[b]{Observatorio Pierre Auger, Av.\ San Mart{\'\i}n Norte 304, 5613 Malarg\"ue, Argentina\\
Full author list: {\rm\url{https://www.auger.org/archive/authors_icrc_2025.html}}}

\emailAdd{spokespersons@auger.org}

\abstract{Rejection of cloud-contaminated data is a complex and important process at the Pierre
Auger Observatory, one which combines information from several sources, including IR cameras, lidars, and satellite imaging. With the deteriorating quality of the IR cameras and challenges
in using other sources, we propose a new method. We use continuous detector monitoring measurements to build a large database of night sky background fluxes for each pixel across
27 telescopes. Using this database, we generate the expected background flux and define
cloud rejection thresholds. Through a straightforward analysis we construct boolean cloud-contamination masks. We demonstrate some results of the analysis, including comparisons with cloud detected using infra-red observations.}

\ConferenceLogo{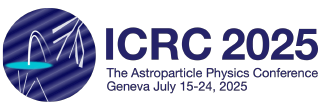}

\FullConference{39th International Cosmic Ray Conference (ICRC2025)\\
 15–24 July 2025\\
Geneva, Switzerland\\}

\begin{document}
\maketitle

\section{Introduction}
The Pierre Auger Observatory uses the atmosphere as a calorimeter to detect ultra-high energy cosmic rays (UHECR).  The Fluorescence Detector (FD) observes the fluorescence light emitted by nitrogen molecules excited by a cosmic ray-induced Extensive Air Shower (EAS).  The Observatory hosts four FD sites on the boundary of a 3000\,km$^2$ array of surface detectors.  A total of 27 fluorescence telescopes view the atmosphere above the array.  Each telescope consists of a 3.8\,m$^2$ entrance aperture containing a near-UV glass filter, a 13\,m$^2$ 
spherical mirror and a 440-pixel camera, each pixel viewing a $1.5^\circ$ diameter section of the sky~\cite{NIM2015}.

Clouds in the atmosphere can obscure the fluorescence light from EAS, resulting in erroneous measurements.  So far, our primary cloud detection instruments have been infra-red (IR) cameras (providing pixel-by-pixel cloud information for the FD cameras), augmented by GOES weather satellites and local lidars~\cite{HarveyICRC2019}. Given that the aging IR cameras and their steering mechanisms have become unreliable, we demonstrate a new method for cloud detection. 

During its operation, the FD runs a continuous calibration process that samples the variance of the baseline trace for every pixel every 30 seconds, and these data are available over the life of the Observatory.  The baseline variance can be converted to a measurement of Night-Sky Background (NSB) photon flux, as described below. The presence of cloud can then be inferred from a sky brightness that is dimmer than expected for a given sidereal time and direction.

\section{Conversion of variances to NSB photon flux}
Due to the AC coupling of the FD photomultiplier tubes (PMTs), a direct measurement of the baseline flux is impossible. It can be obtained indirectly through the statistical analysis of the PMT current fluctuations. Sampling of the signal variance for each pixel is done every 30\,s for a duration of 6.5\,ms in bins of 100\,ns.  Conversion of the baseline variance data to NSB photon fluxes requires the absolute calibration of the each FD pixel, as well as characteristics of the electronics for that channel~\cite{SegretoICRC2011}. The NSB photon flux (in 365\,nm-equivalent photons/m$^2$/sr/s) is given by
\eq{
\phi_\gamma = \frac{[\sigma^2_\text{ADC}]^\text{NSB} \, K_V \, C_\text{FD}}{A \, \Omega \, \Delta t}
\label{phi_gamma}
}
where $[\sigma^2_\text{ADC}]^\text{NSB}$ is the baseline NSB variance (in units of ADC$^2$), $C_\text{FD}$ is the pixel calibration constant 
(in units of 365nm-equivalent photons at the telescope aperture per ADC count), $A$ is the effective telescope area, $\Omega$ is the pixel solid-angle, and $\Delta t$ is the time-bin width.
The factor $K_V$ 
is measured nightly through a calibration process for every pixel. 
It is defined as the mean current during the calibration light pulse divided by the variance in the signal during the pulse, 
\eq{
K_\text{V} = \frac{I_\text{ADC}}{\sigma^2_\text{ADC}}  = \frac{1}{G(1+V_G)}\frac{F_S}{2F}
\label{KV}
}
where the current is measured in ADC counts, and the variance in squared-ADC counts.  While an experimental measurement, $K_V$ can be interpreted in terms of PMT/electronics parameters as shown the right-hand part of \cref{KV}~\cite{Kleifges2002}.  There, $G$ is the combined PMT/electronics gain (ADC counts per photo-electron), $V_G$ is the PMT gain variance (the contribution to the signal variance due to inefficiencies in photo-electron collection within the PMT), $F_S$ is the digitization sampling frequency, and $F$ is the cut-off frequency of the low-pass filter ahead of the digitizer.

\section{Contributions to the Night Sky Background}

\begin{figure}[!h]
    \centering
    \def\w{0.49}
    \includegraphics[width=\w\textwidth]{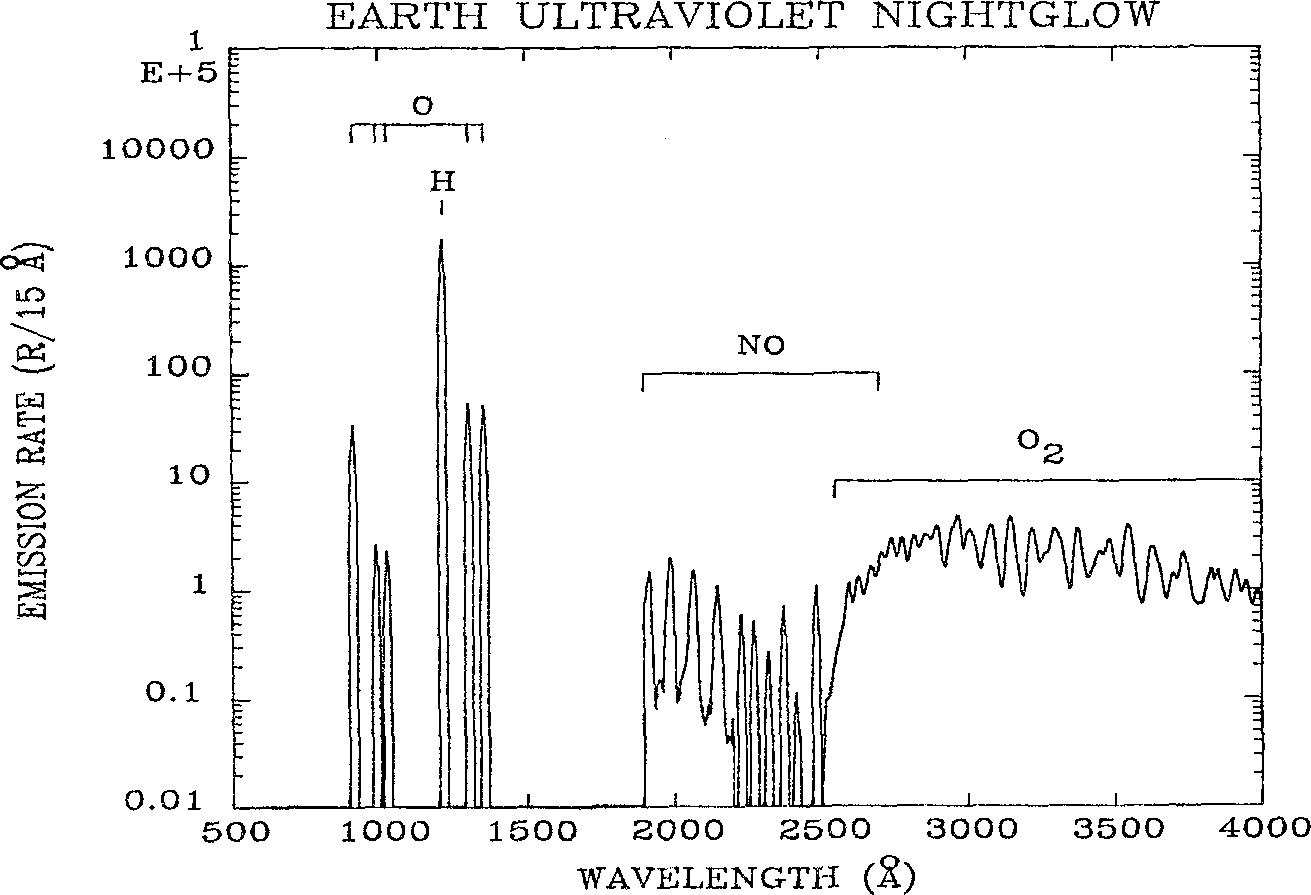} 
    \hfill
    \includegraphics[width=\w\textwidth]{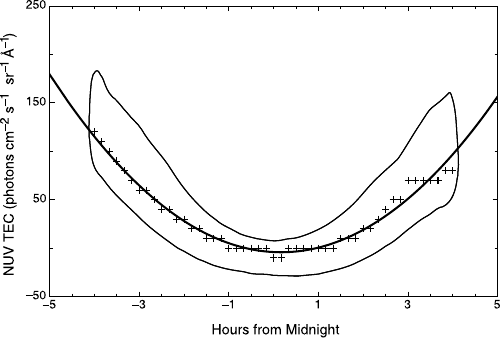}
    \caption{Near-UV nightglow.  \emph{Left:}  the spectrum from Ref.~\cite{Meier1991} showing the Herzberg I molecular oxygen bands in the near-UV.  \emph{Right:} the temporal variation of the near-UV nightglow with respect to local midnight, showing the mean strength and the one-sigma variation~\cite{Murthy2014}.}
    \label{f:nightglow}
\end{figure}

Typical values of the NSB flux are $\sim 5$ photons/m$^2$/deg$^2$/(100\,ns), or approximately 40 photons per 100\,ns integration period for a pixel, but with obvious departures due to the presence of bright stars or intervening cloud.  The sources of NSB in the relevant wavelength band ($\sim 300$ to $440$\,nm) include stars and airglow.  While the star field is predictable, the airglow is highly variable. Night-time airglow (sometimes called nightglow) occurs when atoms or molecules that were photo-ionised or dissociated during the day participate in reactions that release photons~\cite{Meier1991}.  
The spectrum is shown in \cref{f:nightglow}~(left).  The intensity of the nightglow lingers in the west for several hours after sunset and appears in the east several hours before sunrise (\cref{f:nightglow}~right), reflecting the availability of dissociated oxygen atoms necessary for the production of the light.

The elevation-angle dependence of star light and airglow is different.  Stars become dimmer at lower elevations due to atmospheric attenuation, while the airglow is stronger at low elevations because the thickness of the emitting layer is enhanced.  In the latter case, the (upper) atmosphere is the source of the emission, not the medium attenuating it.  

The expected NSB image for a clear sky will depend on the positions of the stars at a particular local sidereal time (LST) and the level of airglow present.  There will be a small dependence on the aerosol conditions.

\section{Determining the threshold for cloud assignment}

In our method, cloud is detected by its dimming effect on the expected night-sky background.  We have produced a set of threshold templates for all FD telescopes that change as a function of LST.  We set thresholds pixel by pixel, an update of our previous work~\cite{JasonMPhil} where we defined only elevation-dependent thresholds for each telescope.  The pixel-based threshold can take into account the very common NSB variations with azimuth at a given elevation.

The variability of the airglow can mean that the clear-sky brightness is not constant in a particular direction, even at the same sidereal time,  but we determine the threshold based on the \emph{lowest} clear-sky brightness at a particular LST.   An extra-bright sky is then interpreted as the absence of cloud. This assumption may break in the presence of the moon, where moonlight can illuminate clouds, making them appear brighter than expected thresholds (see discussion in \cref{performance}).

\begin{figure}
    \centering
    \def\h{0.37}
    \includegraphics[height=\h\textwidth]{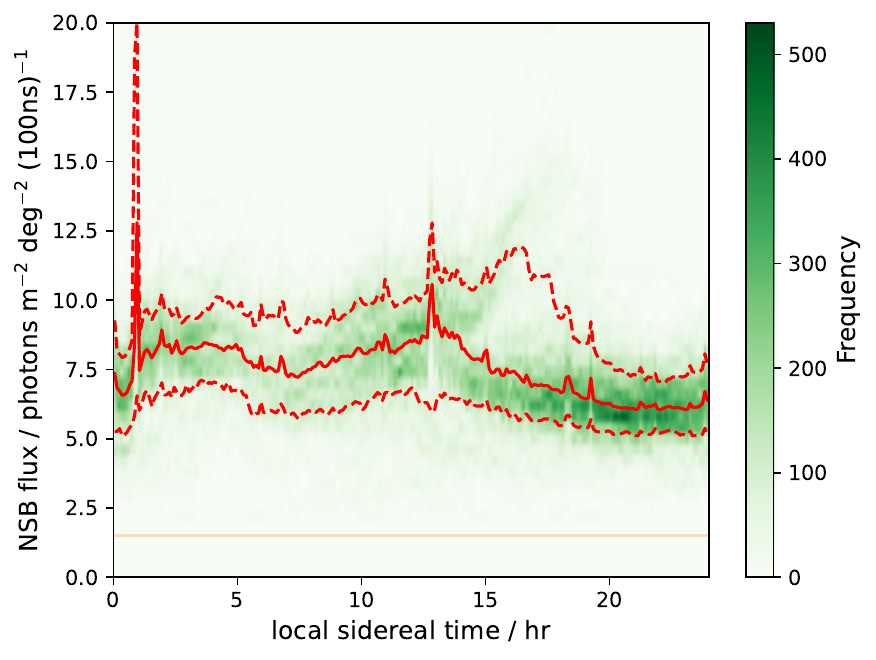} 
    \hfill
    \includegraphics[height=\h\textwidth]{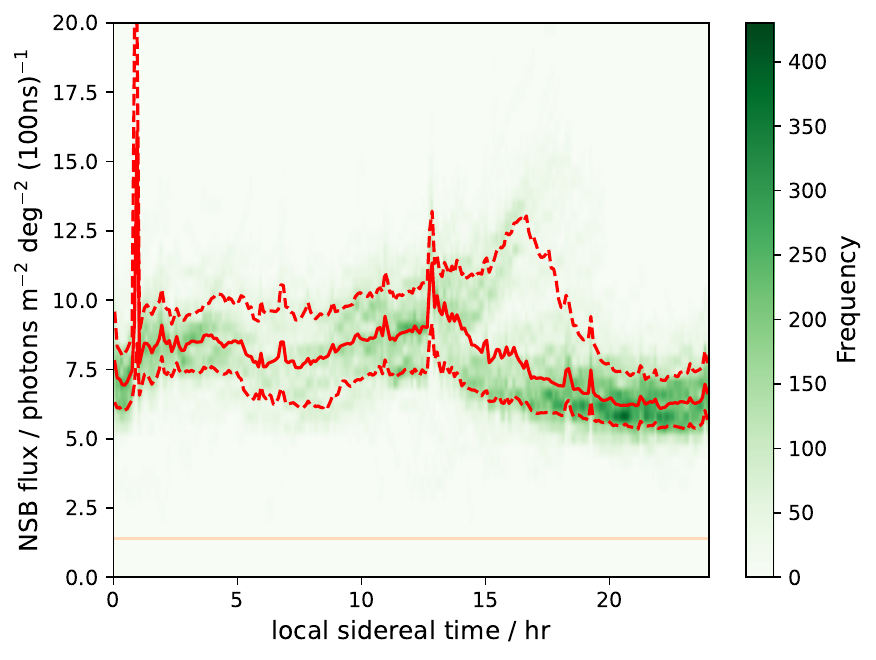}
    \caption{2D histograms of NSB flux as a function of local sidereal time for five years of data (2017-21) from Coihueco FD site telescope 4 pixel 210, a pixel close to the centre of the telescope field of view. \emph{Left:} for all observations with the moon and sun at least $18^\circ$ below the horizon.  The red lines show the median, and the 16th and 84th percentiles, of the NSB flux as a function of LST.  The solid horizontal line at a low flux level represents the typical closed-shutter signal. \emph{Right:} with an additional requirement of minimum hourly cloud-base height $>8000$\,m, determined by our bi-static lidars.}
    \label{f:thresholdVstime}
\end{figure}

As mentioned, NSB measurements are taken for every pixel every 30 seconds during operation.  In \cref{f:thresholdVstime}~(left) we show a summary of all measurements for a particular pixel over a five-year period, in the form of a 2D histogram of NSB flux vs.\ sidereal time.  A cut has been applied to ensure that the sun and moon are at least $18^\circ$ below the horizon, and the red lines show (from the bottom) the 16th percentile, the median, and the 84th percentile of the NSB flux.  One can see obvious variation with LST caused by a changing star-field, as well as variations at any given LST caused by airglow activity and cloud.  For the right-hand plot in the same figure, we have applied an additional cut based on measurements from bi-static lidars of the minimum cloud-base height in each hour. A close comparison of the two figures reveals fewer low-flux measurements in the right-hand plot, especially at 13 hours, where a commonly occurring cloud obscures a passing star, changing the 16th percentile curve. This cut produces a better estimate of the dimmest clear-sky flux.

The choice of the 8000\,m cloud base height cut is a compromise between eliminating cloudy nights and retaining a sufficiently large data set to construct a valid 16th percentile threshold. We assume that some thin, high-altitude cloud contributes to NSB flux measurements below the 16th percentile thresholds. However, any effects of high, thin cloud on EAS measurements are considered minimal, as the bulk of shower development occurs in the lower 5\,km of the atmosphere.

\section{Producing the cloud mask}

Our choice for a cloud threshold is the 16th percentile NSB flux. We note that a stable estimation of this threshold requires several years of data shown in \cref{f:thresholdVstime}, particularly in the LST range observed during short summer nights and in the range with strong airglow variability.  Once defined, the threshold is the minimum NSB flux to be associated with a clear sky for a pixel at a given LST.  We define a \emph{graded cloud index} as a number between 0 and 1,
\eq{
C_\text{graded}=
\begin{cases}
0 & \text{; if flux $F \geq F_\text{16}$ (clear sky)} 
\\
(F_\text{16}-F)/(F_\text{16}-F_\text{shutter})& \text{; otherwise (cloudy)}
\end{cases},
\label{e:gradedIndex}
}
given a pixel flux $F$, a cloud-threshold $F_\text{16}$ and the closed-shutter flux $F_\text{shutter}$. The latter is non-zero because of PMT and electronic noise (see the horizontal lines in \cref{f:thresholdVstime}).  To produce graded cloud masks, smoothing is performed - the graded index of a pixel is averaged with the indices of the pixel's nearest neighbours.  

\begin{figure}
    \centering
    \def\w{0.49}
    \includegraphics[width=\w\textwidth]{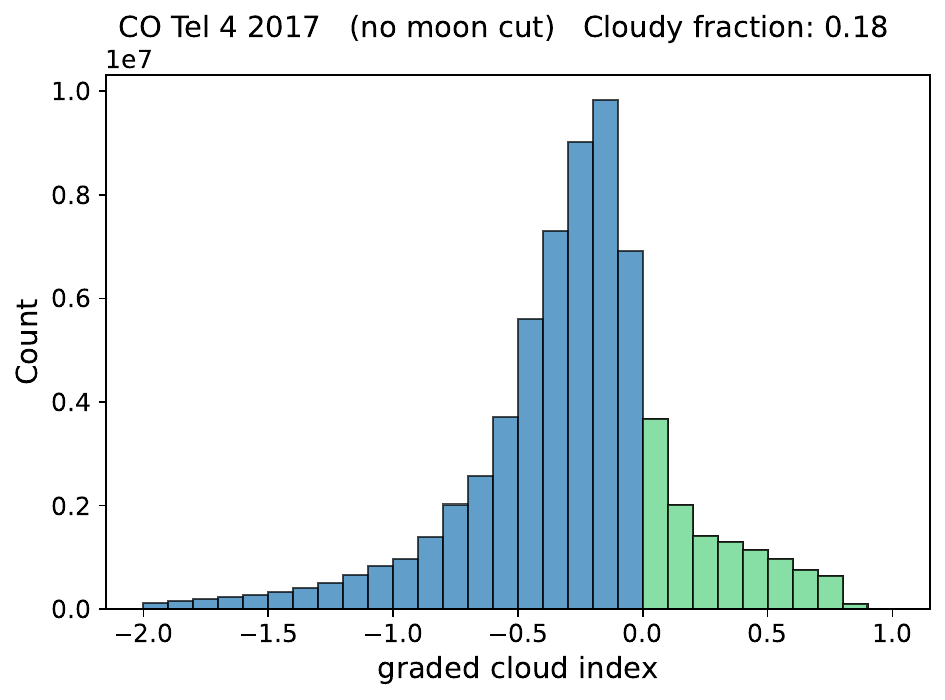} 
    \hfill
    \includegraphics[width=\w\textwidth]{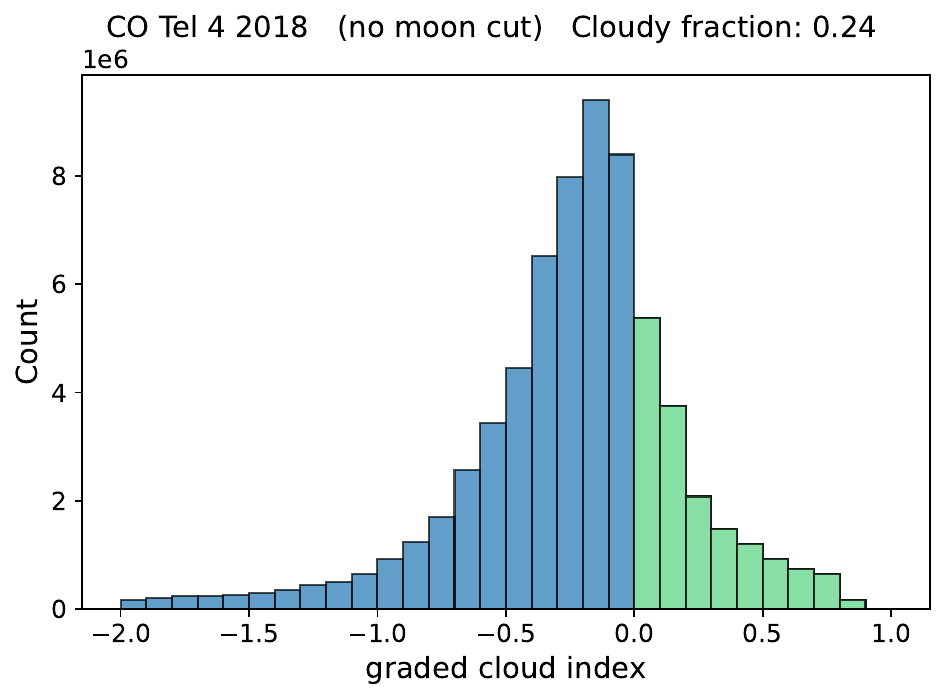}
    \caption{Examples of the full distribution of the graded cloud index.  The negative entries refer to pixel fluxes \emph{brighter} than the 16th percentile NSB flux, and for these $C_\text{graded}$ is set to zero.  The positive values represent cloudy pixels.  The cloudy fraction in a plot title refers to the fraction of histogram entries that are positive i.e.\ cloudy. \emph{Left:} Coihueco Telescope 4 for 2017. \emph{Right:} The same telescope in 2018.}
    \label{f:gradedIndex}
\end{figure}

The distributions of the graded cloud index function $(F_\text{16}-F)/(F_\text{16}-F_\text{shutter})$ are shown in \cref{f:gradedIndex} for a telescope in 2017 and 2018. Negative values represent a clear sky and are set to 0 in cloud masks.  Thus in \cref{f:gradedIndex}, blue-shaded regions correspond to a clear sky, and green areas represent cloud.  
Finally, we define a binary cloud index $C_\text{binary}$. We replace all values of $C_\text{graded} > 0$ with unity.  Some smoothing of $C_\text{binary}$ is also done within a telescope by flipping the state of isolated cloudy or clear pixels.

\begin{figure}
    \centering
    \def\w{1}
    \includegraphics[width=\w\textwidth]{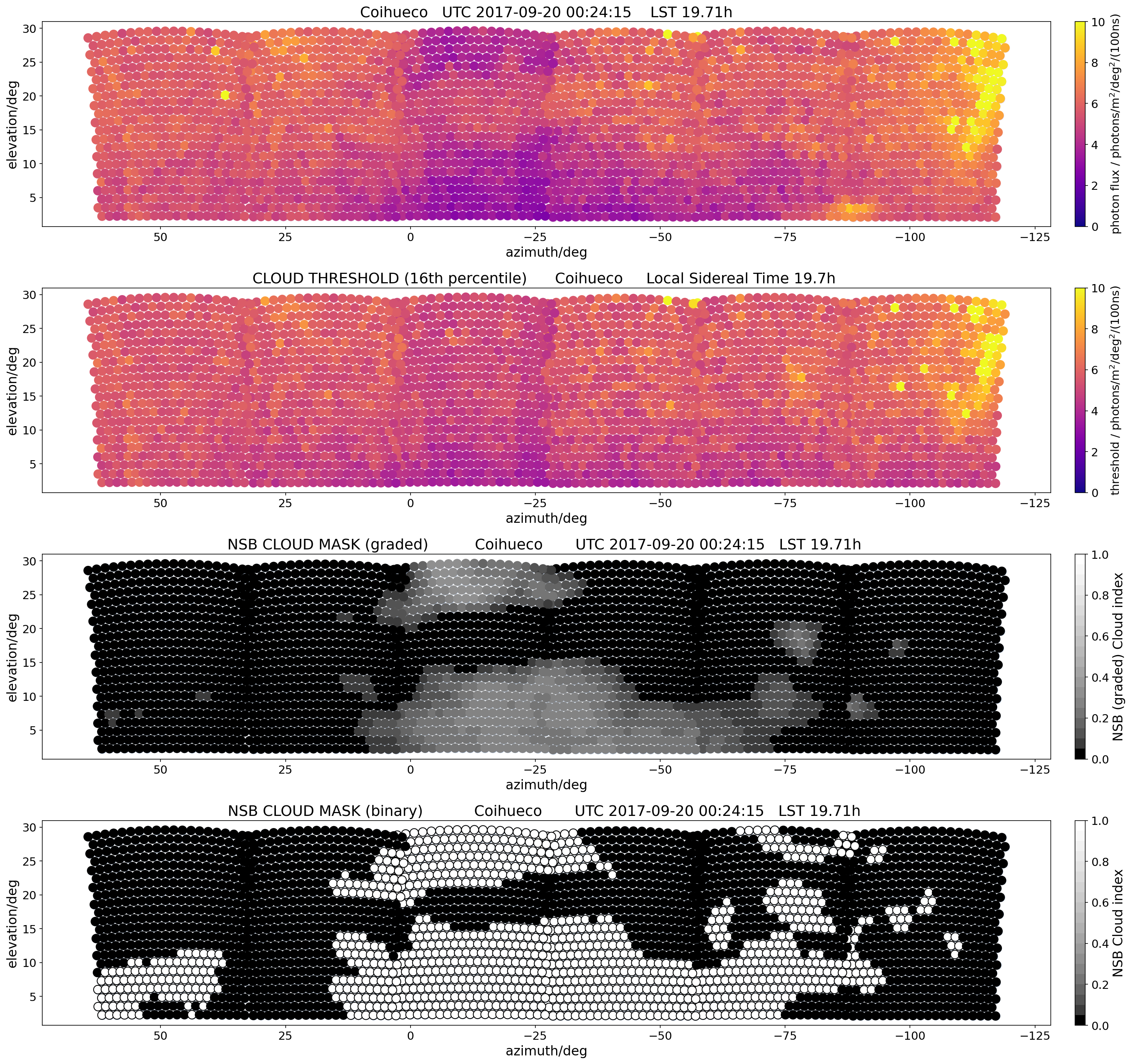} 
    \caption{A snapshot of the sky from FD Coihueco on 20 September, 2017.  The four panels show the NSB flux for all pixels at 00:24 UTC, a local sidereal time of 19.71\,h; the cloud thresholds for all pixels for the 0.1\,h sidereal time interval centered on 19.7\,h;  the graded cloud mask $C_\text{graded}$; and the binary cloud mask $C_\text{binary}$.  At this time, the next sunrise was 10 hours away, and the sun and moon elevations were $23.2^\circ$ and $27.2^\circ$ below the horizon, respectively.  An animation of these plots over a period of 2.5\,hr is available~\cite{Four-panel movie}.}
    \label{f:four-panel}
\end{figure}

\section{Application and Performance}
\label{performance}

An example of a particular snapshot of the sky for Coihueco is shown in \cref{f:four-panel}~(top panel).  The second panel shows a map of cloud thresholds (16th percentile) for the sidereal time of the observation, and the third panel shows the derived graded cloud indices for that snapshot, including the smoothing described above.  The fourth panel shows the binary cloud indices, after smoothing to remove isolated clear/cloudy pixels.  A time-series set of images of this type are available in an animation~\cite{Four-panel movie}, where one can clearly see cloud moving through the field of view of the telescopes.

We have examined the performance of the method.  As a simple cross-check, \cref{f:MonthlyMeans} shows the monthly average of cloudiness in one telescope at each of the four FD sites in 2022.  The four telescopes are at widely dispersed locations and view distinct directions and different NSB fluxes, and yet there is a consistency in the cloudiness detected.  Perfect consistency is not expected because the different telescopes view different directions in varying topography.  Similar results are obtained for other years, and different telescope choices.

The advantage of detecting cloud via the NSB is that the cloud information is available at all times of FD operation.  While similar in function, the infra-red cloud cameras situated at each FD site~\cite{HarveyICRC2019} were unable to provide complete coverage in time due to hardware and software issues.  Additionally, the analysis of IR data was complicated by the non-uniformity in sensitivity across an IR camera's wide field of view (vignetting), and by its sensitivity to atmospheric water vapour in addition to the observationally-important cloud.  In \cref{f:NSB_IRa} we show a comparison of cloudiness viewed by a telescope in a 5-hour period according to the NSB and the IR analyses.  (This period of time corresponds the plots in \cref{f:four-panel} and the animation in \cite{Four-panel movie}.  A brief rain-shower at $t\sim 180$\,min triggered an operator-shutdown of the FD for about an hour).  The agreement is good, but not perfect, with the NSB method seeing somewhat more cloud.  

\noindent
\begin{minipage}[t]{0.48\textwidth}
  \begin{figure}[H]
    \centering
    \includegraphics[width=\linewidth]{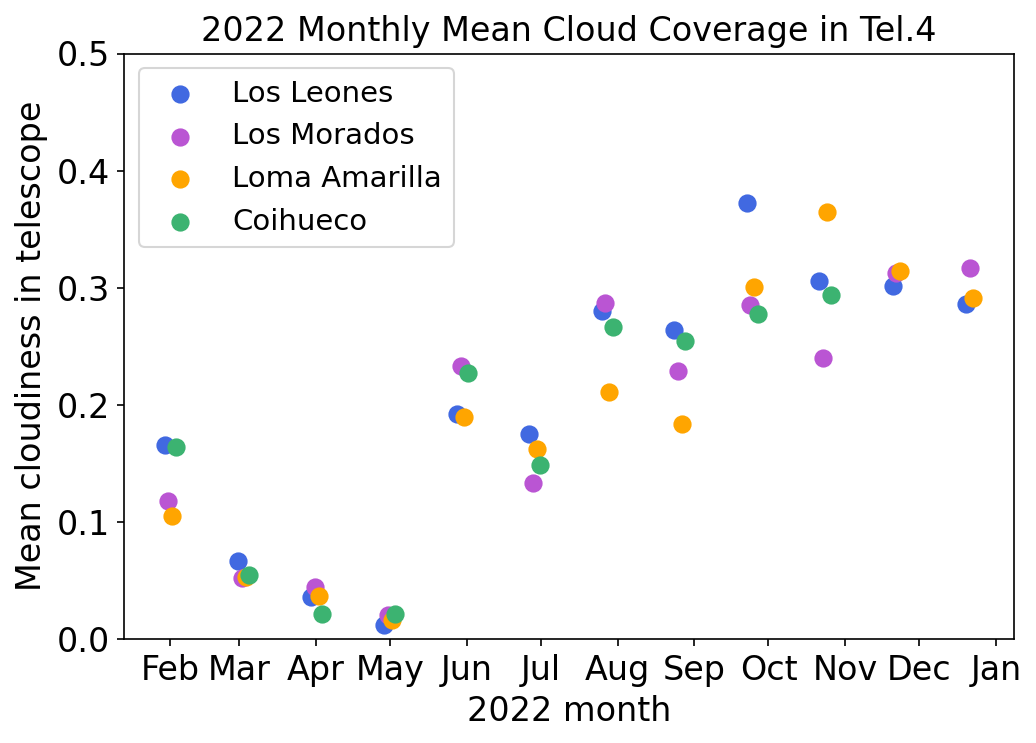}
    \caption{Cloudiness as a function of time for one telescope at each site in 2022.  For each telescope we show the mean value of the binary cloud index across the telescope, averaged over the lunar month.}
    \label{f:MonthlyMeans}
  \end{figure}
\end{minipage}%
\hfill
\begin{minipage}[t]{0.48\textwidth}
  \begin{figure}[H]
    \centering
    \includegraphics[width=\linewidth]{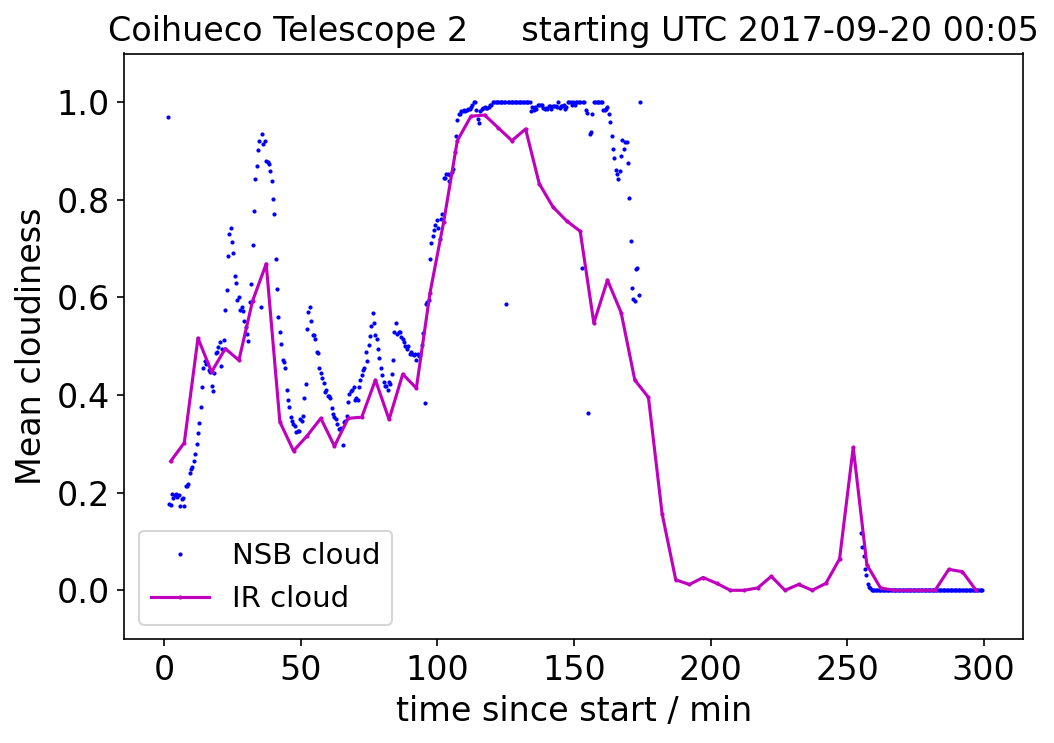}
    \caption{Mean cloud across a telescope in a 5-hour period, comparing infra-red and NSB methods.  The images in \cref{f:four-panel} correspond to $t=19$\,min on this plot, and the animation in \cite{Four-panel movie} covers the first 2.5\,hr.}  
    \label{f:NSB_IRa}
  \end{figure}
\end{minipage}

\noindent
\begin{minipage}[t]{0.48\textwidth}
  \begin{figure}[H] 
    \centering
    \includegraphics[width=\linewidth]{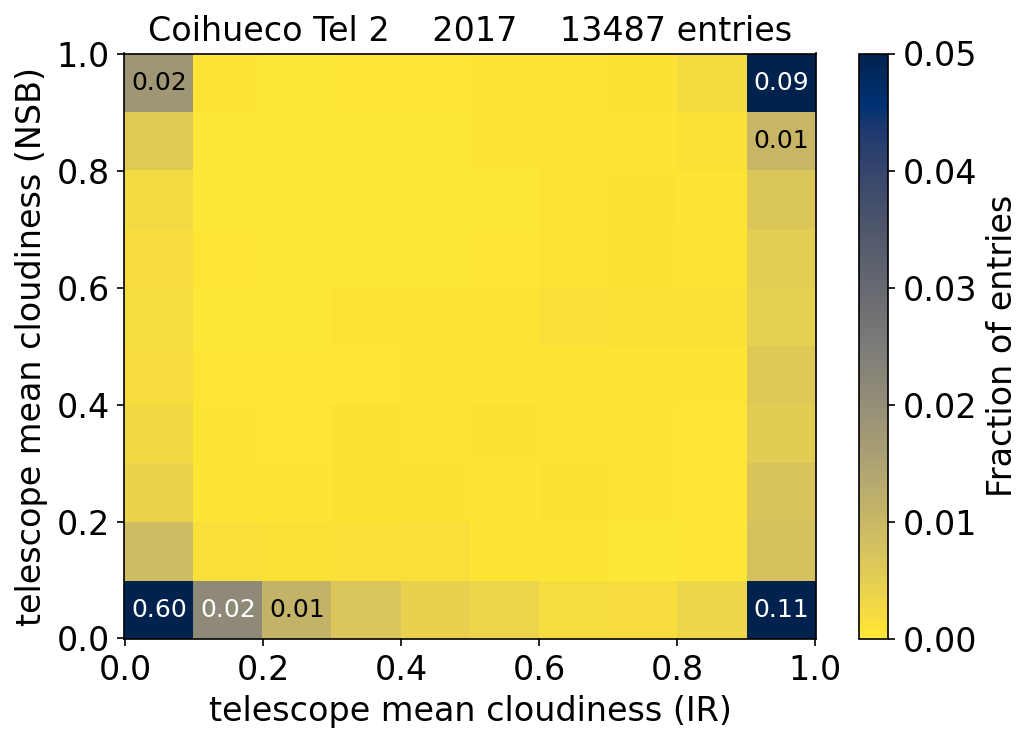}
    \caption{For every 5 minutes of operation during 2017, we show a comparison of the cloudiness in Telescope 2 at Coihueco as measured by the two methods, NSB and IR.  Bins on the 2D histogram are labelled if the fraction of entries is 0.01 or more.}
    \label{f:NSB_IRb}
  \end{figure}
\end{minipage}%
\hfill
\begin{minipage}[t]{0.48\textwidth}
  \begin{figure}[H] 
    \centering
    \includegraphics[width=\linewidth]{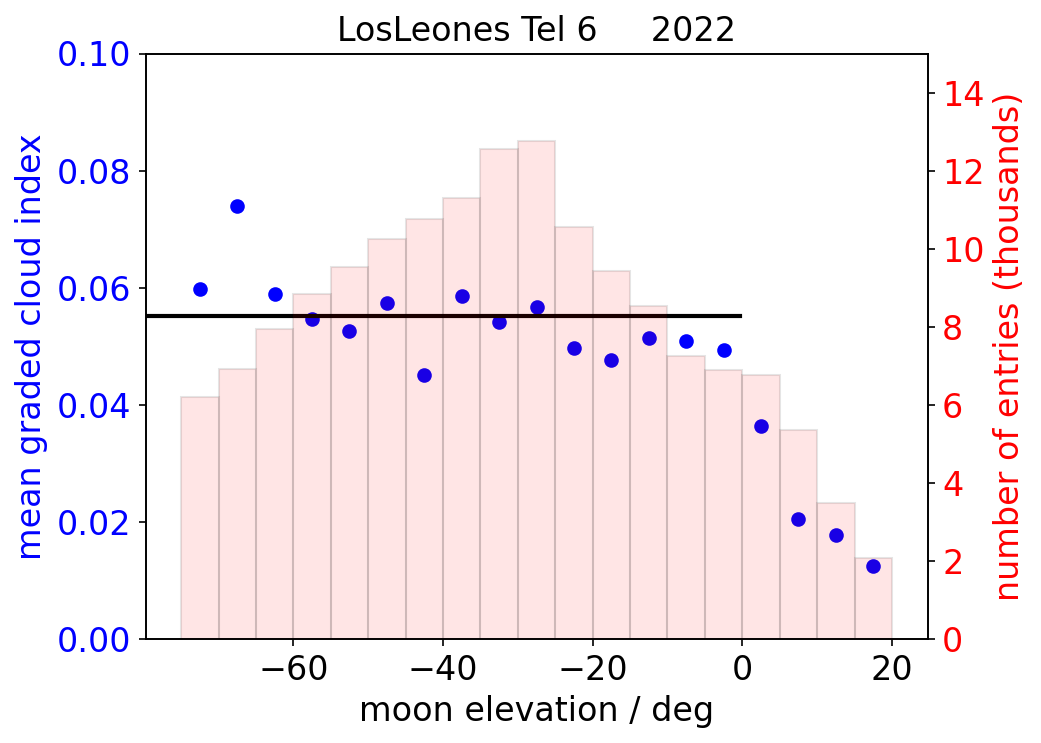}
    \caption{The effect of the moon. In blue, a measure of cloudiness (mean of the graded cloud index, see \cref{e:gradedIndex}) as a function of moon elevation.  The histogram and the right-hand scale shows the distribution of observations with moon elevation.}
    \label{f:moon_effect}
  \end{figure}
\end{minipage}

\vspace{5mm}
A comparison for the same telescope over a full year is shown in \cref{f:NSB_IRb}.  We note that independent of method, the distribution of cloudiness is rather bimodal, either clear or overcast.  This is expected, as periods of partly-cloudy conditions are short lived.  Also note the bin at the bottom-right, where 11\% of images are identified as cloudy by the IR camera, but clear by the NSB method.  Through comparisons with bi-static lidar measurements, we have established that in these cases the IR cameras are viewing high ($> 9$\,km asl) cirrus cloud very unlikely to affect air shower observations.  Thus in terms of relevant cloud, we see a high degree of agreement between the NSB and infra-red methods.

Finally, we point out one small, and expected, deficiency in the NSB method.  \cref{f:moon_effect} shows evidence for a decrease in the efficiency of cloud detection under a crescent moon.   This is expected for some relative positions of the moon, cloud and detector when cloud might be illuminated by the moon, making it brighter than the clear-sky expectation.    The figure shows a rather stable cloudiness as a function of moon elevation when the moon is below the horizon, and a smaller mean cloudiness when the moon has risen, indicating a reduced efficiency of cloud detection.  We find from this example, and generally from other telescopes and years, that the effectiveness of cloud detection drops to 30\% of its dark-sky efficiency during the 15\% of FD operation time conducted with a crescent moon.  This is an acceptable level of impact, and during the times of reduced efficiency, more weight is given to other cloud detection methods such as lidars and satellite observations.

\section{Conclusions}

At the Pierre Auger Observatory, cloud detection via IR cameras has been replaced by a method using night-sky background light, measurements of which have been made routinely during the life of the FD telescopes.  These cloud detections are combined with information from other instruments, including the height of the cloud base, to identify air-shower events affected by cloud~\cite{HarveyICRC2019}.  The NSB method takes advantage of existing well-calibrated FD telescopes designed for shower detection, mitigating the need for separate instruments with their associated maintenance. Although the method is somewhat less sensitive to cloud during the 15\% of FD operations when moonlight is present, this must be weighed against the significant inefficiencies suffered with the previous technique using IR cameras, caused by hardware failures and the insensitivity of the IR technique in times of high humidity\footnote{IR camera information was available for 80\% of air-shower events in years when their operation was smooth, but the long-term efficiency was 67\%.}. 

In general, the NSB cloud detection method is an excellent replacement for our IR cloud cameras.  For the type of cloud relevant to EAS observations, we find equally good performance in the new method, with added advantages of increased sampling (every 30 seconds vs. 5 minutes for the IR cameras) and the ability to fill gaps in the previous cloud database using historical NSB data.

\clearpage
\section*{The Pierre Auger Collaboration}

{\footnotesize\setlength{\baselineskip}{10pt}
\noindent
\begin{wrapfigure}[11]{l}{0.12\linewidth}
\vspace{-4pt}
\includegraphics[width=0.98\linewidth]{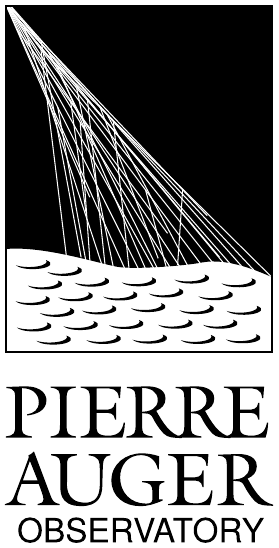}
\end{wrapfigure}
\begin{sloppypar}\noindent
\input{AuthorList/latex_authorlist_authors}
\end{sloppypar}
\begin{center}
\end{center}

\vspace{1ex}
\input{AuthorList/latex_authorlist_institutions}

\input{AuthorList/acknowledgments}
}

\end{document}

%% file: AuthorList/latex_authorlist_authors.tex
A.~Abdul Halim$^{13}$,
P.~Abreu$^{70}$,
M.~Aglietta$^{53,51}$,
I.~Allekotte$^{1}$,
K.~Almeida Cheminant$^{78,77}$,
A.~Almela$^{7,12}$,
R.~Aloisio$^{44,45}$,
J.~Alvarez-Mu\~niz$^{76}$,
A.~Ambrosone$^{44}$,
J.~Ammerman Yebra$^{76}$
% ...

%% file: AuthorList/latex_authorlist_institutions.tex
\begin{description}[labelsep=0.2em,align=right,labelwidth=0.7em,labelindent=0em,leftmargin=2em,noitemsep,before={\renewcommand\makelabel[1]{##1 }}]
\item[$^{1}$] Centro At\'omico Bariloche and Instituto Balseiro (CNEA-UNCuyo-CONICET), San Carlos de Bariloche, Argentina
\item[$^{2}$] Departamento de F\'\i{}sica and Departamento de Ciencias de la Atm\'osfera y los Oc\'eanos, FCEyN, Universidad de Buenos Aires and CONICET, Buenos Aires, Argentina
\item[$^{3}$] IFLP, Universidad Nacional de La Plata and CONICET, La Plata, Argentina
\item[$^{4}$] Instituto de Astronom\'\i{}a y F\'\i{}sica del Espacio (IAFE, CONICET-UBA), Buenos Aires, Argentina
\item[$^{5}$] Instituto de F\'\i{}sica de Rosario (IFIR) -- CONICET/U.N.R.\ and Facultad de Ciencias Bioqu\'\i{}micas y Farmac\'euticas U.N.R., Rosario, Argentina
\item[$^{6}$] Instituto de Tecnolog\'\i{}as en Detecci\'on y Astropart\'\i{}culas (CNEA, CONICET, UNSAM), and Universidad Tecnol\'ogica Nacional -- Facultad Regional Mendoza (CONICET/CNEA), Mendoza, Argentina
\item[$^{7}$] Instituto de Tecnolog\'\i{}as en Detecci\'on y Astropart\'\i{}culas (CNEA, CONICET, UNSAM), Buenos Aires, Argentina
\item[$^{8}$] International Center of Advanced Studies and Instituto de Ciencias F\'\i{}sicas, ECyT-UNSAM and CONICET, Campus Miguelete -- San Mart\'\i{}n, Buenos Aires, Argentina
\item[$^{9}$] Laboratorio Atm\'osfera -- Departamento de Investigaciones en L\'aseres y sus Aplicaciones -- UNIDEF (CITEDEF-CONICET), Argentina
\item[$^{10}$] Observatorio Pierre Auger, Malarg\"ue, Argentina
\item[$^{11}$] Observatorio Pierre Auger and Comisi\'on Nacional de Energ\'\i{}a At\'omica, Malarg\"ue, Argentina
\item[$^{12}$] Universidad Tecnol\'ogica Nacional -- Facultad Regional Buenos Aires, Buenos Aires, Argentina
\item[$^{13}$] University of Adelaide, Adelaide, S.A., Australia
\item[$^{14}$] Universit\'e Libre de Bruxelles (ULB), Brussels, Belgium
\item[$^{15}$] Vrije Universiteit Brussels, Brussels, Belgium
\item[$^{16}$] Centro Brasileiro de Pesquisas Fisicas, Rio de Janeiro, RJ, Brazil
\item[$^{17}$] Centro Federal de Educa\c{c}\~ao Tecnol\'ogica Celso Suckow da Fonseca, Petropolis, Brazil
\item[$^{18}$] Instituto Federal de Educa\c{c}\~ao, Ci\^encia e Tecnologia do Rio de Janeiro (IFRJ), Brazil
\item[$^{19}$] Universidade de S\~ao Paulo, Escola de Engenharia de Lorena, Lorena, SP, Brazil
\item[$^{20}$] Universidade de S\~ao Paulo, Instituto de F\'\i{}sica de S\~ao Carlos, S\~ao Carlos, SP, Brazil
\item[$^{21}$] Universidade de S\~ao Paulo, Instituto de F\'\i{}sica, S\~ao Paulo, SP, Brazil
\item[$^{22}$] Universidade Estadual de Campinas (UNICAMP), IFGW, Campinas, SP, Brazil
\item[$^{23}$] Universidade Estadual de Feira de Santana, Feira de Santana, Brazil
\item[$^{24}$] Universidade Federal de Campina Grande, Centro de Ciencias e Tecnologia, Campina Grande, Brazil
\item[$^{25}$] Universidade Federal do ABC, Santo Andr\'e, SP, Brazil
\item[$^{26}$] Universidade Federal do Paran\'a, Setor Palotina, Palotina, Brazil
\item[$^{27}$] Universidade Federal do Rio de Janeiro, Instituto de F\'\i{}sica, Rio de Janeiro, RJ, Brazil
\item[$^{28}$] Universidad de Medell\'\i{}n, Medell\'\i{}n, Colombia
\item[$^{29}$] Universidad Industrial de Santander, Bucaramanga, Colombia
\item[$^{30}$] Charles University, Faculty of Mathematics and Physics, Institute of Particle and Nuclear Physics, Prague, Czech Republic
\item[$^{31}$] Institute of Physics of the Czech Academy of Sciences, Prague, Czech Republic
\item[$^{32}$] Palacky University, Olomouc, Czech Republic
\item[$^{33}$] CNRS/IN2P3, IJCLab, Universit\'e Paris-Saclay, Orsay, France
\item[$^{34}$] Laboratoire de Physique Nucl\'eaire et de Hautes Energies (LPNHE), Sorbonne Universit\'e, Universit\'e de Paris, CNRS-IN2P3, Paris, France
\item[$^{35}$] Univ.\ Grenoble Alpes, CNRS, Grenoble Institute of Engineering Univ.\ Grenoble Alpes, LPSC-IN2P3, 38000 Grenoble, France
\item[$^{36}$] Universit\'e Paris-Saclay, CNRS/IN2P3, IJCLab, Orsay, France
\item[$^{37}$] Bergische Universit\"at Wuppertal, Department of Physics, Wuppertal, Germany
\item[$^{38}$] Karlsruhe Institute of Technology (KIT), Institute for Experimental Particle Physics, Karlsruhe, Germany
\item[$^{39}$] Karlsruhe Institute of Technology (KIT), Institut f\"ur Prozessdatenverarbeitung und Elektronik, Karlsruhe, Germany
\item[$^{40}$] Karlsruhe Institute of Technology (KIT), Institute for Astroparticle Physics, Karlsruhe, Germany
\item[$^{41}$] RWTH Aachen University, III.\ Physikalisches Institut A, Aachen, Germany
\item[$^{42}$] Universit\"at Hamburg, II.\ Institut f\"ur Theoretische Physik, Hamburg, Germany
\item[$^{43}$] Universit\"at Siegen, Department Physik -- Experimentelle Teilchenphysik, Siegen, Germany
\item[$^{44}$] Gran Sasso Science Institute, L'Aquila, Italy
\item[$^{45}$] INFN Laboratori Nazionali del Gran Sasso, Assergi (L'Aquila), Italy
\item[$^{46}$] INFN, Sezione di Catania, Catania, Italy
\item[$^{47}$] INFN, Sezione di Lecce, Lecce, Italy
\item[$^{48}$] INFN, Sezione di Milano, Milano, Italy
\item[$^{49}$] INFN, Sezione di Napoli, Napoli, Italy
\item[$^{50}$] INFN, Sezione di Roma ``Tor Vergata'', Roma, Italy
\item[$^{51}$] INFN, Sezione di Torino, Torino, Italy
\item[$^{52}$] Istituto di Astrofisica Spaziale e Fisica Cosmica di Palermo (INAF), Palermo, Italy
\item[$^{53}$] Osservatorio Astrofisico di Torino (INAF), Torino, Italy
\item[$^{54}$] Politecnico di Milano, Dipartimento di Scienze e Tecnologie Aerospaziali , Milano, Italy
\item[$^{55}$] Universit\`a del Salento, Dipartimento di Matematica e Fisica ``E.\ De Giorgi'', Lecce, Italy
\item[$^{56}$] Universit\`a dell'Aquila, Dipartimento di Scienze Fisiche e Chimiche, L'Aquila, Italy
\item[$^{57}$] Universit\`a di Catania, Dipartimento di Fisica e Astronomia ``Ettore Majorana``, Catania, Italy
\item[$^{58}$] Universit\`a di Milano, Dipartimento di Fisica, Milano, Italy
\item[$^{59}$] Universit\`a di Napoli ``Federico II'', Dipartimento di Fisica ``Ettore Pancini'', Napoli, Italy
\item[$^{60}$] Universit\`a di Palermo, Dipartimento di Fisica e Chimica ''E.\ Segr\`e'', Palermo, Italy
\item[$^{61}$] Universit\`a di Roma ``Tor Vergata'', Dipartimento di Fisica, Roma, Italy
\item[$^{62}$] Universit\`a Torino, Dipartimento di Fisica, Torino, Italy
\item[$^{63}$] Benem\'erita Universidad Aut\'onoma de Puebla, Puebla, M\'exico
\item[$^{64}$] Unidad Profesional Interdisciplinaria en Ingenier\'\i{}a y Tecnolog\'\i{}as Avanzadas del Instituto Polit\'ecnico Nacional (UPIITA-IPN), M\'exico, D.F., M\'exico
\item[$^{65}$] Universidad Aut\'onoma de Chiapas, Tuxtla Guti\'errez, Chiapas, M\'exico
\item[$^{66}$] Universidad Michoacana de San Nicol\'as de Hidalgo, Morelia, Michoac\'an, M\'exico
\item[$^{67}$] Universidad Nacional Aut\'onoma de M\'exico, M\'exico, D.F., M\'exico
\item[$^{68}$] Institute of Nuclear Physics PAN, Krakow, Poland
\item[$^{69}$] University of \L{}\'od\'z, Faculty of High-Energy Astrophysics,\L{}\'od\'z, Poland
\item[$^{70}$] Laborat\'orio de Instrumenta\c{c}\~ao e F\'\i{}sica Experimental de Part\'\i{}culas -- LIP and Instituto Superior T\'ecnico -- IST, Universidade de Lisboa -- UL, Lisboa, Portugal
\item[$^{71}$] ``Horia Hulubei'' National Institute for Physics and Nuclear Engineering, Bucharest-Magurele, Romania
\item[$^{72}$] Institute of Space Science, Bucharest-Magurele, Romania
\item[$^{73}$] Center for Astrophysics and Cosmology (CAC), University of Nova Gorica, Nova Gorica, Slovenia
\item[$^{74}$] Experimental Particle Physics Department, J.\ Stefan Institute, Ljubljana, Slovenia
\item[$^{75}$] Universidad de Granada and C.A.F.P.E., Granada, Spain
\item[$^{76}$] Instituto Galego de F\'\i{}sica de Altas Enerx\'\i{}as (IGFAE), Universidade de Santiago de Compostela, Santiago de Compostela, Spain
\item[$^{77}$] IMAPP, Radboud University Nijmegen, Nijmegen, The Netherlands
\item[$^{78}$] Nationaal Instituut voor Kernfysica en Hoge Energie Fysica (NIKHEF), Science Park, Amsterdam, The Netherlands
\item[$^{79}$] Stichting Astronomisch Onderzoek in Nederland (ASTRON), Dwingeloo, The Netherlands
\item[$^{80}$] Universiteit van Amsterdam, Faculty of Science, Amsterdam, The Netherlands
\item[$^{81}$] Case Western Reserve University, Cleveland, OH, USA
\item[$^{82}$] Colorado School of Mines, Golden, CO, USA
\item[$^{83}$] Department of Physics and Astronomy, Lehman College, City University of New York, Bronx, NY, USA
\item[$^{84}$] Michigan Technological University, Houghton, MI, USA
\item[$^{85}$] New York University, New York, NY, USA
\item[$^{86}$] University of Chicago, Enrico Fermi Institute, Chicago, IL, USA
\item[$^{87}$] University of Delaware, Department of Physics and Astronomy, Bartol Research Institute, Newark, DE, USA
\item[] -----
\item[$^{a}$] Max-Planck-Institut f\"ur Radioastronomie, Bonn, Germany
\item[$^{b}$] also at Kapteyn Institute, University of Groningen, Groningen, The Netherlands
\item[$^{c}$] School of Physics and Astronomy, University of Leeds, Leeds, United Kingdom
\item[$^{d}$] Fermi National Accelerator Laboratory, Fermilab, Batavia, IL, USA
\item[$^{e}$] Pennsylvania State University, University Park, PA, USA
\item[$^{f}$] Colorado State University, Fort Collins, CO, USA
\item[$^{g}$] Louisiana State University, Baton Rouge, LA, USA
\item[$^{h}$] now at Graduate School of Science, Osaka Metropolitan University, Osaka, Japan
\item[$^{i}$] Institut universitaire de France (IUF), France
\item[$^{j}$] now at Technische Universit\"at Dortmund and Ruhr-Universit\"at Bochum, Dortmund and Bochum, Germany
\end{description}

%% file: AuthorList/acknowledgments.tex
\section*{Acknowledgments}

\begin{sloppypar}
The successful installation, commissioning, and operation of the Pierre
Auger Observatory would not have been possible without the strong
commitment and effort from the technical and administrative staff in
Malarg\"ue. We are very grateful to the following agencies and
organizations for financial support:
\end{sloppypar}

\begin{sloppypar}
Argentina -- Comisi\'on Nacional de Energ\'\i{}a At\'omica; Agencia Nacional de
Promoci\'on Cient\'\i{}fica y Tecnol\'ogica (ANPCyT); Consejo Nacional de
Investigaciones Cient\'\i{}ficas y T\'ecnicas (CONICET); Gobierno de la
Provincia de Mendoza; Municipalidad de Malarg\"ue; NDM Holdings and Valle
Las Le\~nas; in gratitude for their continuing cooperation over land
access; Australia -- the Australian Research Council; Belgium -- Fonds
de la Recherche Scientifique (FNRS); Research Foundation Flanders (FWO),
Marie Curie Action of the European Union Grant No.~101107047; Brazil --
Conselho Nacional de Desenvolvimento Cient\'\i{}fico e Tecnol\'ogico (CNPq);
Financiadora de Estudos e Projetos (FINEP); Funda\c{c}\~ao de Amparo \`a
Pesquisa do Estado de Rio de Janeiro (FAPERJ); S\~ao Paulo Research
Foundation (FAPESP) Grants No.~2019/10151-2, No.~2010/07359-6 and
No.~1999/05404-3; Minist\'erio da Ci\^encia, Tecnologia, Inova\c{c}\~oes e
Comunica\c{c}\~oes (MCTIC); Czech Republic -- GACR 24-13049S, CAS LQ100102401,
MEYS LM2023032, CZ.02.1.01/0.0/0.0/16{\textunderscore}013/0001402,
CZ.02.1.01/0.0/0.0/18{\textunderscore}046/0016010 and
CZ.02.1.01/0.0/0.0/17{\textunderscore}049/0008422 and CZ.02.01.01/00/22{\textunderscore}008/0004632;
France -- Centre de Calcul IN2P3/CNRS; Centre National de la Recherche
Scientifique (CNRS); Conseil R\'egional Ile-de-France; D\'epartement
Physique Nucl\'eaire et Corpusculaire (PNC-IN2P3/CNRS); D\'epartement
Sciences de l'Univers (SDU-INSU/CNRS); Institut Lagrange de Paris (ILP)
Grant No.~LABEX ANR-10-LABX-63 within the Investissements d'Avenir
Programme Grant No.~ANR-11-IDEX-0004-02; Germany -- Bundesministerium
f\"ur Bildung und Forschung (BMBF); Deutsche Forschungsgemeinschaft (DFG);
Finanzministerium Baden-W\"urttemberg; Helmholtz Alliance for
Astroparticle Physics (HAP); Helmholtz-Gemeinschaft Deutscher
Forschungszentren (HGF); Ministerium f\"ur Kultur und Wissenschaft des
Landes Nordrhein-Westfalen; Ministerium f\"ur Wissenschaft, Forschung und
Kunst des Landes Baden-W\"urttemberg; Italy -- Istituto Nazionale di
Fisica Nucleare (INFN); Istituto Nazionale di Astrofisica (INAF);
Ministero dell'Universit\`a e della Ricerca (MUR); CETEMPS Center of
Excellence; Ministero degli Affari Esteri (MAE), ICSC Centro Nazionale
di Ricerca in High Performance Computing, Big Data and Quantum
Computing, funded by European Union NextGenerationEU, reference code
CN{\textunderscore}00000013; M\'exico -- Consejo Nacional de Ciencia y Tecnolog\'\i{}a
(CONACYT) No.~167733; Universidad Nacional Aut\'onoma de M\'exico (UNAM);
PAPIIT DGAPA-UNAM; The Netherlands -- Ministry of Education, Culture and
Science; Netherlands Organisation for Scientific Research (NWO); Dutch
national e-infrastructure with the support of SURF Cooperative; Poland
-- Ministry of Education and Science, grants No.~DIR/WK/2018/11 and
2022/WK/12; National Science Centre, grants No.~2016/22/M/ST9/00198,
2016/23/B/ST9/01635, 2020/39/B/ST9/01398, and 2022/45/B/ST9/02163;
Portugal -- Portuguese national funds and FEDER funds within Programa
Operacional Factores de Competitividade through Funda\c{c}\~ao para a Ci\^encia
e a Tecnologia (COMPETE); Romania -- Ministry of Research, Innovation
and Digitization, CNCS-UEFISCDI, contract no.~30N/2023 under Romanian
National Core Program LAPLAS VII, grant no.~PN 23 21 01 02 and project
number PN-III-P1-1.1-TE-2021-0924/TE57/2022, within PNCDI III; Slovenia
-- Slovenian Research Agency, grants P1-0031, P1-0385, I0-0033, N1-0111;
Spain -- Ministerio de Ciencia e Innovaci\'on/Agencia Estatal de
Investigaci\'on (PID2019-105544GB-I00, PID2022-140510NB-I00 and
RYC2019-027017-I), Xunta de Galicia (CIGUS Network of Research Centers,
Consolidaci\'on 2021 GRC GI-2033, ED431C-2021/22 and ED431F-2022/15),
Junta de Andaluc\'\i{}a (SOMM17/6104/UGR and P18-FR-4314), and the European
Union (Marie Sklodowska-Curie 101065027 and ERDF); USA -- Department of
Energy, Contracts No.~DE-AC02-07CH11359, No.~DE-FR02-04ER41300,
No.~DE-FG02-99ER41107 and No.~DE-SC0011689; National Science Foundation,
Grant No.~0450696, and NSF-2013199; The Grainger Foundation; Marie
Curie-IRSES/EPLANET; European Particle Physics Latin American Network;
and UNESCO.
\end{sloppypar}